\documentclass[runningheads]{llncs}
\usepackage{graphicx}
\usepackage{todonotes}
\usepackage[bottom]{footmisc}
\usepackage{booktabs}
\usepackage{multirow}
\usepackage{amssymb}
\DeclareTextFontCommand{\judgment}{\bfseries\itshape}

\newcommand\eugene[1]{{#1}}
\newcommand\doug[1]{{#1}}
\newcommand\jim[1]{{#1}}
\newcommand\dawn[1]{{#1}}

\usepackage{floatrow}
\begin{document}
\title{HC4: A New Suite of Test Collections \\ for Ad Hoc CLIR}
\titlerunning{HC4: A New Suite of Test Collections for Ad Hoc CLIR}
\author{Dawn Lawrie\inst{1}\orcidID{0000-0001-7347-7086} \and
James Mayfield\inst{1}\orcidID{0000-0003-3866-3013} \and \\
Douglas W. Oard\inst{1,2}\orcidID{0000-0002-1696-0407} \and \\
Eugene Yang\inst{1}\orcidID{0000-0002-0051-1535}}

\authorrunning{D. Lawrie et al.}

\institute{
HLTCOE, 
Johns Hopkins University, Baltimore MD 21211, USA \\
\email{\{lawrie,mayfield,eugene.yang\}@jhu.edu}\\
\and
University of Maryland, College Park, College Park MD 20742, USA\\
\email{oard@umd.edu}}

\maketitle              %
\begin{abstract}

HC4 is a new suite of test collections for ad hoc Cross-Language Information Retrieval (CLIR),
with Common Crawl News documents in Chinese, Persian, and Russian,
topics in English and in the document languages,
and graded relevance judgments.
New test collections are needed because existing CLIR test collections built using pooling of traditional CLIR runs
have systematic gaps in their relevance judgments when used to evaluate neural CLIR methods.
The HC4 collections contain 60 topics and about half a million documents for each of Chinese and Persian,
and \dawn{54 topics} and five million documents for Russian. 
Active learning was used to determine which documents to annotate after being seeded using interactive search and judgment.  
\doug{Documents were judged on a three-grade relevance scale.}
This paper describes the design and construction of the new test collections
and provides baseline results for demonstrating their utility for evaluating systems. 

\end{abstract}
\section{Introduction}

Ad hoc Cross-Language Information Retrieval (CLIR) has been studied for decades.
Yet until the advent of high-quality machine translation,
the usefulness of CLIR has been limited.
Easy access to inexpensive or free machine translation has altered this landscape.
If one can find a document of interest in a language one cannot read,
machine translation is now often sufficient to make the majority of the document's content accessible.
Thus, the breadth of the audience for CLIR has increased dramatically in a short period of time.

\jim{As machine translation has increased the usefulness of CLIR,
recently introduced deep neural methods} have improved 
\doug{ranking quality}~\cite{bonab2020training,nair2020combining,yarmohammadi2019robust,zhang2019improving,zhao2019weakly}.
By and large, these 
techniques appear to provide a large jump in the quality of CLIR output.
Yet the evidence for these improvements is based on small, dated test collections~\cite{FerroP09,ferro2019information,MajumderMPBMPMS10,SchaubleS97,Smeaton96}.
Problems with existing collections include:
\begin{itemize}
    \item Some CLIR test collections are no longer available from any standard source.
    \item They are typically small, often 100,000 or fewer documents, and some have few known relevant documents per topic.
    \item Judgment pools were retrieved using older systems.
    New neural systems are thus more likely to systematically identify relevant unjudged documents~\cite{thakur2021beir,voorhees2021coopetition,zhang2021mr}.
    \item Many of the early test collections have only binary judgments.
\end{itemize}
The increased importance of CLIR thus
argues for the creation of new ad hoc CLIR collections that ameliorate these problems.
A new CLIR collection should contain a large number of recent documents in a standard encoding,
with distribution rights that foster broad use,
sufficient numbers of relevant documents per topic to allow systems to be distinguished,
and graded relevance 
\doug{judgments}.

To this end, we have created HC4\footnote{HC4 can be downloaded from \texttt{\url{https://github.com/hltcoe/HC4}}.}
-- the HLTCOE Common Crawl CLIR Collection.
In addition to addressing the shortcomings described above \eugene{and facilitating evaluations of new CLIR systems},
this suite of collections has a few unique aspects.
First, to mimic 
\doug{well contextualized}
search sessions, topics are generally inspired by events in the news and written from the 
perspective of a knowledgeable searcher
familiar with the background information
on the event. %
Each topic is associated with a date,
and in most cases the topic is linked to Wikipedia page text
written immediately prior to that date, generally contemporaneous with the event.
This page serves as a proxy for a report that might have written by a searcher prior to their search,
reflecting their knowledge at that time.
It is included in the collection to enable exploration of contextual search.
Second, to maximize recall in the judged set, instead of pooling, active learning identified the documents to be judged~\cite{hical}.
\jim{This approach reduces judgment bias toward any specific automated retrieval system. }

\section{Related Work}

The first CLIR test collection was created for Salton's seminal work on CLIR in 1970,
in which English queries were manually translated into German~\cite{salton1970automatic}.
Relevance judgments were exhaustively created for those queries for several hundred abstracts in both languages.
In 1995, the first instance of a large-scale CLIR test collection in which documents were selected for assessment using pooling
translated Spanish queries from the Fourth Text Retrieval Conference's (TREC-4) Spanish test collection into English for CLIR experimentation~\cite{DavisD95}.
The next year, TREC organizers provided standard English versions of queries for Spanish and Chinese collections~\cite{Smeaton96}.
The following year, CLIR became the explicit focus of a TREC track, with collections in German, French, and Italian;
that track continued for three years~\cite{SchaubleS97}.
One enduring contribution from this early work was recognition that to be representative of actual use,
translations of topic fields in a test collection should not be made word-by-word,
but rather should be re-expressions fluently written in the 
\doug{query}
language.

With the start of the NACSIS Test Collection Information Retrieval (NTCIR) evaluations in Japan in 1999~\cite{sakai2021evaluating},
the Cross-Language Evaluation Forum (CLEF) in Europe in 2000~\cite{ferro2019information},
and the Forum for Information Retrieval Evaluation (FIRE) in India in 2008~\cite{MajumderMPBMPMS10},
the center of gravity of CLIR evaluation moved away from TREC.
Over time, the research in each of these venues has become more specialized,
so although CLIR tasks continue,
the last large-scale CLIR test collection for ad hoc search of news that was produced in any of the world's four major information retrieval shared-task evaluation venues %
was created in 2009 for Persian~\cite{FerroP09}.
The decline in test collection production largely reflected a relative stasis in CLIR research,
which peaked around the turn of the century and subsequently tailed off.
Perhaps the best explanation for the decline is that the field had,
by the end of the first decade of the twenty-first century,
largely exhausted the potential of the statistical alignment techniques for parallel text that had commanded the attention of researchers in that period.  

One consequence of this hiatus is that older test collections do not always age gracefully.
As Lin et al.~point out,
``Since many innovations work differently than techniques that came before,
old evaluation instruments may not be capable of accurately quantifying effectiveness improvements associated with later techniques''~\cite{lin2021pretrained}.
The key issue here is that in large test collections, relevance judgments are necessarily sparse.
TREC introduced pooling as a way to decide which (typically several hundred) documents should be judged for relevance to each topic,
with the remaining documents remaining unjudged.
Pools were constructed by merging highly ranked documents from a diverse range of fully automated systems,
including some of the best systems of the time,
\jim{sometimes augmented by documents found using interactive search.}
Zobel found, using evaluation measures that treat unjudged documents as not relevant,
that relevance judgments on such pools result in system comparisons not markedly biased against other systems constructed using similar technology that had not contributed to the pools~\cite{zobel1998reliable}.
Contemporaneously, Voorhees found that comparisons between systems were generally insensitive to substituting judgments from one assessor for those of another~\cite{DBLP:journals/ipm/Voorhees00a}. 
A subsequent line of work found that some newly designed evaluation measures produced system comparisons robust to random ablation of those pools~\cite{DBLP:conf/sigir/BuckleyV04,DBLP:journals/tois/MoffatZ08,DBLP:journals/ir/SakaiK08,DBLP:conf/cikm/YilmazA06}.
However, these conclusions do not necessarily hold when new technology finds relevant documents that were not found by earlier methods,
as can be the case for neural retrieval methods~\cite{lin2021pretrained}.
In such cases, three approaches might be tried:
\begin{enumerate}
    \item 
     Re-pool and rejudge an older collection, or create a new collection over newer content using pooling.
    \item 
    Select documents to be judged in a manner relatively insensitive to the search technology of the day,
    without necessarily judging all relevant documents.
    \item 
    Use an approach that simply does a better job of finding most of the relevant documents,
    thus reducing the risk of bias towards any class of system.
\end{enumerate}

We used the third of these approaches to select documents for judgment in HC4.
Specifically, we used the HiCAL system~\cite{CormackZGASGRG19} to identify documents for judgment using active learning.
HiCAL was originally developed to support Technology Assisted Review (TAR) in E-Discovery,
where the goal is to identify the largest practical set of relevant documents at a reasonable cost~\cite{oard2013information,cormack2014evaluation,baron2016perspectives,yang2021cost}.
Similar approaches have been used to evaluate recall-oriented search in the TREC Total Recall and Precision Medicine tracks~\cite{totalrecall2015,totalrecall2016,clef2019ehealth-tar}.
The key idea in HiCAL is to train an initial classifier using a small set of relevance judgments,
and then to use active learning with relevance sampling to identify additional documents for review.
As Lewis found, relevance sampling can be more effective than the uncertainty sampling approach that is more commonly used with active learning
when the prevalence of relevant documents in the collection being searched is low~\cite{DBLP:journals/sigir/Lewis95a}.
This low prevalence of relevant documents is often a design goal for information retrieval test collections,
both because many real information retrieval tasks exhibit low relevance prevalence,
and because (absent an oracle that could fairly sample undiscovered relevant documents)
accurately estimating recall requires reasonably complete annotation of the relevant set.
One concern that might arise with HiCAL is that if the document space is bifurcated,
with little vocabulary overlap between two or more sets of relevant documents,
then HiCAL could get stuck in a local optimum,
exploiting one part of the document space well but missing relevant documents in another.
Experience suggests that this can happen, but that such cases are rare.\footnote{Personal communication with Gordon Cormack.}
In particular, we expect such cases to be exceptionally rare in the news stories on which our HC4 test collections are built,
since journalists typically go out of their way to contextualize the information that they present. 

Early TREC CLIR test collections all included binary relevance judgments,
but the introduction of the Discounted Cumulative Gain (DCG) measure in 2000~\cite{DBLP:conf/sigir/JarvelinK00},
and the subsequent broad adoption of Normalized DCG (nDCG),
increased the demand for relevance judgments with more than two relevance grades
(e.g., highly relevant, somewhat relevant, and not relevant).
Some of the early CLIR work with graded relevance judgments first binarized those judgments
(e.g., either by treating highly and somewhat relevant as relevant,
or by treating only highly relevant as relevant)~\cite{DBLP:conf/ntcir/KandoKNEKH99}.
However, Sakai has noted that using graded relevance in this way can rank systems differently
than would more nuanced approaches that award partial credit for finding partially relevant documents~\cite{sakai2021evaluating}.
\jim{In our baseline runs, we report nDCG using the graded relevance judgments,
then binarize those judgments to report Mean Average Precision (MAP)} \dawn{by treating highly and somewhat relevant as relevant.}

\section{Collection Development Methodology}

We adopted several design principles to create HC4.
First, to develop a multilingual document collection that was easy to distribute,
we chose the Common Crawl News Collection as the basis for the suite of collections.
We applied automatic language identification to determine the language of each document.\footnote{\url{https://github.com/bsolomon1124/pycld3}}
We then assembled Chinese, Persian, and Russian documents from August 2016 to August 2019
into ostensibly\footnote{Language 
\doug{ID}
failure caused some documents in each set to be of the wrong language.} monolingual document sets.
Finally, we automatically identified and eliminated duplicate documents.

The second design principle was to create topics that model the interests of a knowledgeable searcher
who writes about world events.
Such topics enable CLIR research that addresses complex information needs that cannot be answered by a few facts.
Key attributes of a knowledgeable searcher include
a \doug{relative} lack of ambiguity in their information need and 
an increased interest in named entities. 
To support this goal, 
we used events reported in the Wikipedia Current Events Portal
(WCEP)\footnote{https://en.wikipedia.org/wiki/Portal:Current\_events}
as our starting point for topic development.
To support exploration of how additional context information could be used to improve retrieval,
each topic was associated with a contemporaneous report.

A third design principle was to include topics with relevant documents in multiple languages.
Once a topic was developed in one language,
it was vetted for possible use with the document sets of other languages.

\subsection{Topic Development}

Starting from an event summary appearing in WCEP,
a topic developer would learn about that event from the English document that was linked to it,
and from additional documents about the event that were automatically identified as part of the WCEP multi-document summarization dataset~\cite{ghalandari2020}.
Topic developers were bilingual, so they could understand how an English topic related to the event being discussed in the news in \dawn{another} language.
After learning about the event, the topic developer searched a non-English collection to find documents about the event.
After reading a few documents in their language,
they were asked to write a sentence or question describing an information need held by the hypothetical knowledgeable searcher. 
They were then asked to write a three-to-five word summary of the sentence.
The summary became the topic title,
and the sentence became the topic description. 
Next, the topic developer would investigate the prevalence of the topic in the collection.
To do this they would issue one or more document-language queries
and judge ten of the resulting documents. 
Topic developers answered two questions about each document:
(1) How relevant is the most important information on the topic in this document?;
and (2) How valuable is the most important information in this document?
Relevance was judged as \judgment{central}, \judgment{tangential}, \judgment{not-relevant}, or \judgment{unable-to-judge}.
The second question was only posed if the answer to the first question was \judgment{central}.
Allowable answers to the second question were \judgment{very-valuable}, \judgment{somewhat-valuable}, and \judgment{not-valuable}.

To develop topics with relevant documents in more than one language,
the title and description,
along with the event that inspired the topic,
were shown to a topic developer for a different language.
The topic developer searched for the presence of the topic in their language.
As with the initial topic development,
ten documents were judged to evaluate whether the document set supported the topic.
Topic developers were allowed to modify the topic,
which sometimes led to vetting the new topic in the initial language.

\subsection{Relevance Judgments}
\label{sec:judgments}

After topic development, some topics were selected for more complete 
assessment. 
The titles and descriptions of selected topics were vetted by a committee
comprising IR researchers and topic developers.
The committee reviewed each topic to ensure that: 
(a) the title and description were mutually consistent and concise; %
(b) titles consisted of three to five non-stopwords; 
(c) descriptions were complete, grammatical sentences with punctuation and correct spelling;
and (d) topics were focused and likely to have a manageable number of relevant documents. %
Corrections were made by having each committee member suggest new phrasing,
then a topic developer selecting a preferred alternative.

Given the impracticality of judging millions of documents, 
and because most documents are not relevant to a given topic,
we followed the common practice of assessing as many relevant documents as possible,
\doug{deferring to the evaluation measure decisions on how unassessed documents should be treated.}
Because we did not build this collection using a shared task,
we did not have diverse systems to contribute to judgment pools. Thus, we could not use {\em pooling}~\cite{webber2010effect,zobel1998reliable}. 
Instead, we used the active learning system HiCAL~\cite{CormackZGASGRG19}, %
to \doug{iteratively} select documents to be judged.
HiCAL builds a classifier based on the known relevant documents using relevance feedback.
As the assessor judges documents,
the classifier is retrained using the new assessments.
To seed HiCAL's classifier,
we used the first ten documents judged during topic development.
Because the relevance assessor is likely not the person who developed the topic, 
and because the topic might have changed during topic vetting,
those documents are re-judged. %
At least one document must be judged relevant to initialize the classifier.

Once assessment was complete,
assessors provided a translation of the title and description fields into the language of the documents,
and briefly explained (in English) how relevance judgments were made;
these explanations were placed in the topic's narrative field. 
In contrast to the narrative in a typical TREC ad hoc collection, which is written prior to judging documents,
these narratives were written after judgments were made;
users of these collections must therefore be careful not to use the narrative field as part of a query on the topic.

Our target time for assessing a single topic was four hours.
We estimated this would allow us to judge about one hundred documents per topic.
According to the designers of HiCAL,\footnote{Personal communication with Gordon Cormack.}
one can reasonably infer that almost all findable relevant documents have been found
if an assessor judges twenty documents in a row as not relevant. 
From this, we estimated that topics with twenty or fewer relevant documents were likely to be fully annotated after viewing 100 documents.
Treating both \judgment{central} and \judgment{tangential} documents as relevant
would have led to more than twenty relevant documents for most selected topics.
Thus, to support topics that went beyond esoteric facts,
we treated only documents deemed central to the topic as relevant.

We established three relevance levels,
defined from the perspective of a user writing a report on the topic:
\begin{description}
\item[\judgment{Very-valuable}] Information in the document would be found in the lead paragraph of a report \doug{that is later} written on the topic.
\item[\judgment{Somewhat-valuable}] The most valuable information in the document would be found in the remainder of such a report.
\item[\judgment{Not-valuable}] Information in the document might be included in a report footnote, or omitted entirely.
\end{description}

\begin{figure}[tb]
\includegraphics[width=\textwidth]{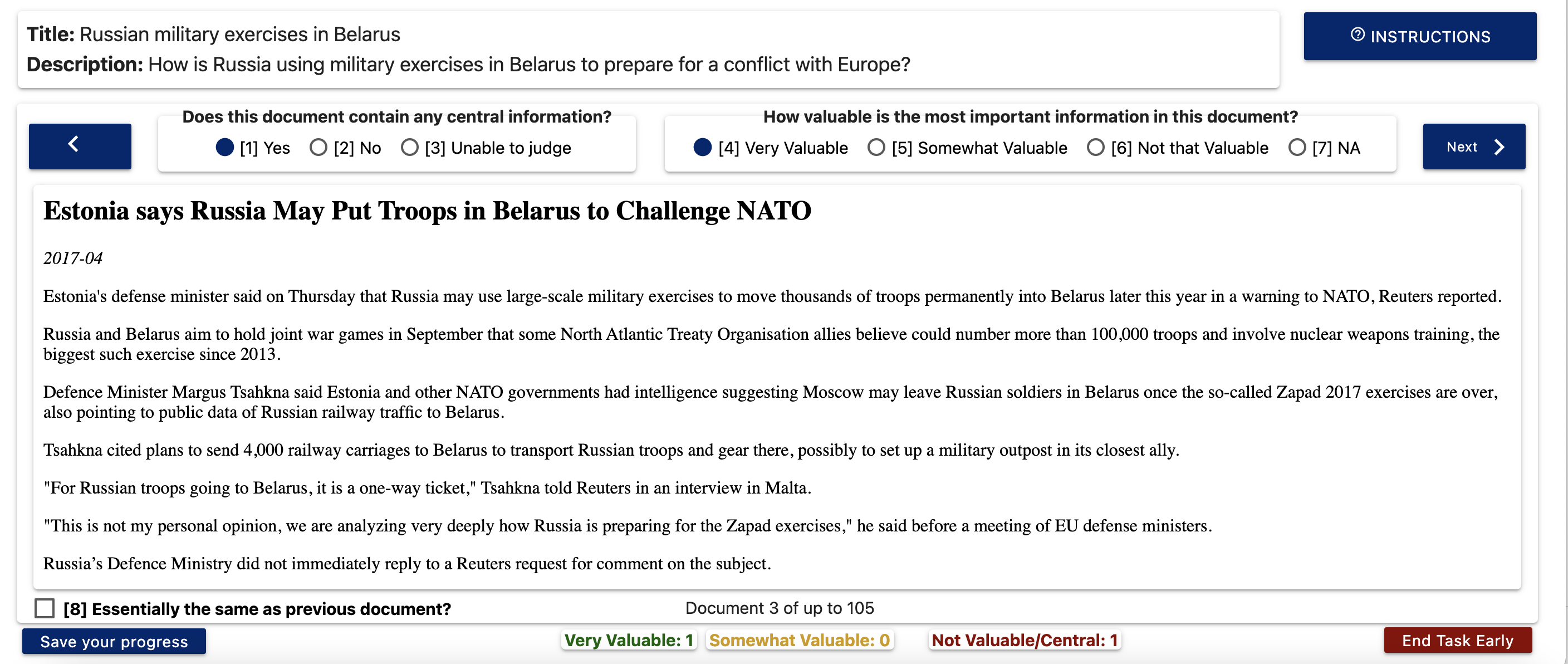}
\caption{Annotation interface for relevance judgments.} \label{fig:phase2}
\end{figure}

To map graded relevance values to the binary relevance required by HiCAL,
documents judged as \judgment{very-valuable} or \judgment{somewhat-valuable} were treated as relevant,
while documents judged \judgment{not-valuable}, and those that were not central to the topic, were considered \judgment{not-relevant}.
The final collection maps the \judgment{not-valuable} category to \judgment{not-relevant}.
This means that a document can mention a topic without being considered relevant to that topic if it
\doug{lacks}
information that would be included in a \doug{future} report.
Because an assessor could judge a topic over multiple days,
assessors took copious notes to 
\doug{foster}
consistency.

To more quickly identify topics too broad to be annotated under our annotation budget,
assessors were instructed to end a task early (eliminating the topic from inclusion in the collection) whenever:
\begin{itemize}
    \item more than five \judgment{very-valuable} or \judgment{somewhat-valuable} documents were found among the first ten assessed;
    \item more than fifteen \judgment{very-valuable} or \judgment{somewhat-valuable} documents were found among the first thirty assessed;
    \item more than forty \judgment{very-valuable} or \judgment{somewhat-valuable} documents were found at any point; or
    \item relevant documents were still being found after assessing 85 or more documents.
\end{itemize}
Once assessment was completed, we dropped any topic with fewer than three relevant documents.
We subsequently sought to refocus dropped topics to ameliorate the problems encountered during assessment;
if this was deemed likely to produce a conforming topic,
the refocused topic was added back into the assessment queue.
Thus, a few similar but not identical topics are present in different languages.

We used the process described above to develop the topics in each of the three languages. 
Figure~\ref{fig:phase2} shows the interface used to annotate the collection.
Key features include:
hot keys to support faster judgment;
next document and previous document navigation;
identification of near-duplicate documents that were not identified during deduplication;
the ability to save progress and return to annotation in another session;
counts of how many documents have been judged in different categories; and
a button to end the annotation early.

\begin{table}[tb]
\caption{Collection statistics.}\label{tab:coll}
\centering
\setlength\tabcolsep{0.4em}
\begin{tabular}{l|cc|cc|cc}
\toprule
 & \multicolumn{2}{c|}{Chinese} & \multicolumn{2}{c|}{Persian} & \multicolumn{2}{c}{Russian} \\
 & Train & Eval& Train & Eval & Train & Eval \\
\midrule
Documents & \multicolumn{2}{c|}{646,305} & \multicolumn{2}{c|}{486,486} & \multicolumn{2}{c}{4,721,064}\\
Topics & 10 & 50 & 10 & 50 & 4 & 50\\
Judged Documents & 466 & 2,751 & 486 & 2,522 & 265 & 2,970 \\
Partially Relevant Documents & 30 & 192 & 46 & 215 & 67 & 411 \\
Highly Relevant Documents & 62 & 282 & 54 & 206 & 12 & 262 \\
\bottomrule
\end{tabular}
\end{table}

\begin{table}[b]
\caption{Multilingual topic counts.}\label{tab:multi}
\centering
\setlength\tabcolsep{0.4em}
\begin{tabular}{l|ccc|c}
\toprule
 & Chinese+Persian & Chinese+Russian & Persian+Russian & All Languages\\ \midrule
Train & 6 & 2 & 2 & 1 \\
Eval& 12 & 14 & 10 & 4 \\
\bottomrule
\end{tabular}
\end{table}

\subsection{Contemporaneous Reports}

Contemporaneous reports are portions of Wikipedia page text written before a particular date. 
Each topic was associated with a date, which either came from the date of the event in WCEP that inspired the topic
or, if after topic development there was no such event, from the earliest relevant document. 
\jim{The assessor was instructed to find the Wikipedia page 
most related to the topic and use the edit history of that page to view it as it appeared on the day before the date listed in the topic. 
The assessor selected text from this page to serve as the contemporaneous report.}
Because of the date restriction, some contemporaneous reports are less closely related to the topic,
since a specific Wikipedia page for the event may not have existed on the day before the event.

\section{Collection Details}

This section introduces collection details, %
discusses the annotation cost 
in terms of time, and reports on inter-assessor agreement. 
Table~\ref{tab:coll} describes the size of the collection in documents and topics, and presents counts
of the number of annotations used in the final collection.  Disjoint subsets of Train and Eval topics are defined to encourage consistent choices by users of the test collections. 
As in most information retrieval collections, the vast majority of the unjudged documents are not relevant.
However, because we used active learning to suggest documents for assessment,
and because of our desire to create topics with relatively few relevant documents,
on average there are only about 50 judged documents per topic. This number ranges from 28 (when no
additional relevant documents were discovered during the second phase) to 112 
documents (when an assessor used the ``Essentially the same" button shown in 
Figure~\ref{fig:phase2}\footnote{This button applies the previous relevance judgment without 
increasing the counter;
it was typically used when several news sources picked up the same story,
but modified it sufficiently to prevent its being automatically labeled as a near duplicate.}).
Some of the topics have judged documents in multiple languages. Table~\ref{tab:multi} displays the number of topics with 
judgments in each pair of languages, and the subset of those with judgments in all three languages. 
While we sought to maximize the number of multilingual topics, we were constrained 
by our annotation budget.

The people who performed topic development and relevance assessment were all bilingual. A majority of them were 
native English speakers, although a few were native speakers in the language of the documents.
While some were proficient in more than two languages, none was
proficient in more than one of Chinese, Persian or Russian. 
Highly fluent topic developers verified that the human translations of topics were expressed fluently in the
non-English language. %

\subsection{Development and Annotation Time}

As a proxy for the cost of creating these test collections,
we report the time spent on topic development and relevance 
\doug{assessment.}
The total time for developing candidate topics, including those not included in the final collection,
is shown in Table~\ref{tab:duration_topic_dev}.
A total of about 570 hours were spent by 30 developers to create the 559 topics in the three languages. 
The median time to develop a topic was about 36 minutes, with an average of about an hour, suggesting a long tail distribution.

\begin{table}[t]
\centering
\caption{Document annotation time in minutes with median of each class and Spearman's $\rho$ correlation between assessment time and the resulting binarized label. }\label{tab:duration_annot}
\setlength\tabcolsep{0.8em}

\begin{tabular}{lc|cc|r}
\toprule
Language  & \# Doc. (Rel / Not) &  Median (Rel / Not) & $\rho$ & Total Time \\
\midrule
Chinese  &   1,094 / 3,863 &   1.33 / 0.75 &    0.1900 &   8,730.65 \\
Persian  &   1,576 / 4,444 &   1.35 / 0.80 &    0.1617 &  11,807.66 \\
Russian  &   2,746 / 5,525 &   0.79 / 0.69 &    0.0584 &  11,561.20 \\
\bottomrule
\end{tabular}
\vspace{-1em}
\end{table}

As mentioned in Section~\ref{sec:judgments}, developed topics were filtered before 
\doug{assessment.}
As shown in Table~\ref{tab:duration_annot},
a total of about 540 hours were spent by 33 assessors.\footnote{We replaced the longest 5\% of assessment times with the median per language,
since these cases likely reflect assessors who left a job unfinished overnight.}
These figures include documents rejudged for quality assurance,
and topics with incomplete assessments. %
The median annotation time per document suggests that relevant documents took longer to judge.
Here, we aggregated \judgment{very-valuable} and \judgment{somewhat-valuable} as relevant, and the remaining categories as not relevant.
Despite this consistent observation across all three languages,
Spearman's $\rho$ suggests only a weak correlation between the judgment time and relevance
due to the long tail distribution shown in Figure~\ref{fig:hist_duration}.
There are more \judgment{not-relevant} documents that took a shorter time to assess, but as we observe in Figure~\ref{fig:hist_duration} the distributions are similar, and the differences are thus not statistically significant by an independent samples t-test.

\begin{figure}[t]
\newfloatcommand{capbtabbox}{table}[][\FBwidth]
\CenterFloatBoxes
\begin{floatrow}
\ttabbox{%
    \setlength\tabcolsep{0.6em}
    \begin{tabular}{l|c|cc}
    \toprule
    Language & Topics & Average & Median \\ 
    \midrule
    Chinese & 240 & 56.49 & 30.57 \\
    Persian & 148 & 52.63 & 36.02 \\ 
    Russian & 181 & 81.60 & 46.58 \\
    \bottomrule
    \end{tabular}
}{%
  \caption{Topic development time in minutes.}\label{tab:duration_topic_dev}
}
\killfloatstyle
\ffigbox{%
    \includegraphics[width=\linewidth]{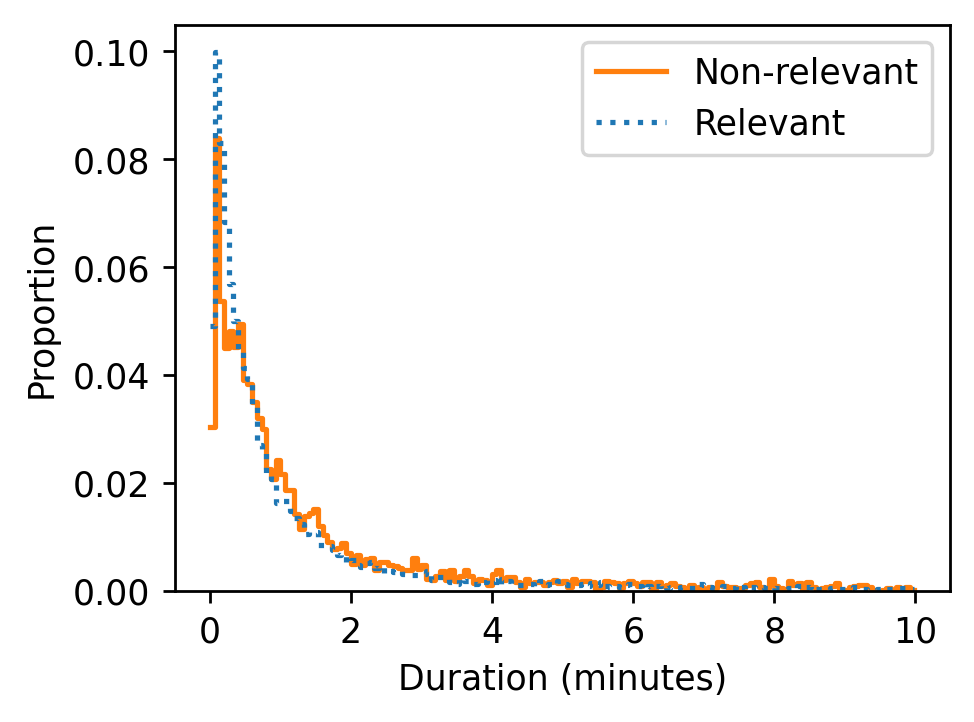}
}{%
    \caption{Document annotation time.}
    \label{fig:hist_duration}
}

\end{floatrow}
\end{figure}

\begin{table}[b]
\def\v{\checkmark}
\caption{Example for intersection and union agreement}\label{tab:agreement_example}
\centering
\setlength\tabcolsep{0.6em}
\begin{tabular}{c|ccccc}
\toprule
Assessor \textbackslash ~Document 
    & D1 & D2 & D3 & D4 & D5 \\
\midrule
A1  & \v &    &    & \v & \v \\
A2  & \v & \v &    & \v &    \\
A3  & \v &    &    & \v & \v \\
\midrule
(I)ntersection / (U)nion
    & I/U&  I &    & I/U& I \\ 
\bottomrule
\end{tabular}
    
\end{table}

\subsection{Inter-assessor Agreement}

\begin{table}[t]

\caption{Inter-assessor agreement on binarized labels. }\label{tab:agreement}

\centering
\setlength\tabcolsep{0.6em}
\begin{tabular}{l|c|cc|cc}
\toprule
 & & \multicolumn{2}{c|}{Intersection} & \multicolumn{2}{c}{Union} \\
Language & \# of topics &  Agreement &  Fleiss' $\kappa$ &  Agreement &  Fleiss' $\kappa$  \\
\midrule
Chinese  &  5 &       0.85 &           0.69 &       0.84 &           0.62 \\
Persian  &  4 &       0.73 &           0.40 &       0.69 &           0.35 \\
Russian  &  3 &       0.69 &           0.33 &       0.69 &           0.33 \\
\bottomrule
\end{tabular}
\vspace{-1em}
\end{table}

Although all topics were assessed by a single assessor for consistency,
several were additionally assessed by one or two other assessors for quality assurance.
In Table~\ref{tab:agreement} we report the raw agreement
(i.e., proportion of the documents all assessors agreed upon)
and the Fleiss' $\kappa$
(i.e., the agreement after chance correction for multiple assessors).
\jim{Because active learning is path-dependent, each assessor judged a somewhat different set of documents;
we thus evaluate agreement on both the intersection and the union of the documents for a complete picture.}
Unjudged documents were considered \judgment{not-relevant} for the union agreements.
Table~\ref{tab:agreement_example} shows an example, where only \texttt{D1} and \texttt{D4} are in the intersection,
judged by all three assessors.
\texttt{D3} was not judged by any assessor, and is thus is not in the union. 

All three languages demonstrate 
\doug{at least}
fair agreement~($\kappa$ between 0.20 and 0.40~\cite{landis1977measurement}),
with Chinese topics having a substantial agreement~($\kappa$ between 0.60 and 0.80), for both the intersection and the union.
The raw agreement 
\doug{indicates}
that 
\doug{69\% to 85\%}
of the judged documents have the same binarized judgments.
The small gap between intersection and union agreements supports our assumption that unjudged documents are not relevant.

\section{Baseline Runs}
To demonstrate the utility of HC4 for evaluating CLIR systems, we report 
retrieval evaluation results for a set of baseline CLIR systems on the Eval sets in Table~\ref{tab:baselines}. Three retrieval approaches, implemented by Patapsco~\cite{patapsco}, human query translation, machine query translation, and machine document translation, use BM25 ($k_1=0.9$, $b=0.4$) with RM3 pseudo relevance feedback on title queries. \eugene{Translation models are trained in-house using the Sockeye toolkit~\cite{hieber2017sockeye}.}

\eugene{
As examples of neural CLIR models, we evaluated \textit{vanilla} reranking models~\cite{macavaney2019cedr} fine-tuned with MS-MARCO-v1~\cite{bajaj2016ms} for at most one epoch with various multi-language pretrained models, including multilingual-BERT~(mBERT)~\cite{devlin2018bert}, XLM-Roberta-large~(XLM-R)~\cite{conneau2019unsupervised}, and infoXLM-large~\cite{chi2020infoxlm}. Model checkpoints were selected by nDCG@100 on HC4 dev sets. Each trained model reranks the top 1000 documents retrieved by the machine query translation BM25 model\footnote{Hence, the input of the reranking models is still English queries with documents in the target language.} in a zero-shot fashion~\cite{colbert-x}.}

\begin{table}
\caption{Baseline results of title queries using BM25 with RM3 on Eval sets, \textit{QT/DT}: query/document translation.
}\label{tab:baselines}
\newcommand{\bb}[1]{\textbf{#1}}
\centering
\setlength\tabcolsep{0.25em}
\begin{tabular}{ll|cc|cc|cc|cc}
\toprule
&     & \multicolumn{2}{c|}{nDCG@100} 
      & \multicolumn{2}{c|}{MAP@100} 
      & \multicolumn{2}{c|}{R@1000} 
      & \multicolumn{2}{c}{Judged@10} \\
Language & Method &     Mean & p-value &    Mean & p-value &    Mean & p-value &    Mean & p-value \\
\midrule
\multicolumn{1}{l}{\multirow{6}{*}{Chinese}} 
& Human QT   &    0.473 &    --  &    0.317 &    --  &    0.825 &    --  &\bb{0.544}&    --  \\
& Machine QT &    0.362 &  0.009 &    0.231 &  0.023 &    0.708 &  0.036 &    0.408 &  0.010 \\
& Machine DT &    0.502 &  0.566 &    0.336 &  0.704 &\bb{0.871}&  0.345 &    0.542 &  0.968 \\
& mBERT      &    0.348 &  0.008 &    0.205 &  0.011 &    0.708 &  0.036 &    0.302 &  0.000 \\
& infoXLM    &\bb{0.541}&  0.155 &\bb{0.369}&  0.297 &    0.708 &  0.036 &    0.504 &  0.428 \\
& XLM-R      &    0.536 &  0.168 &    0.368 &  0.318 &    0.708 &  0.036 &    0.500 &  0.376 \\
\midrule  
\multicolumn{1}{l}{\multirow{6}{*}{Persian}} 
& Human QT   &    0.428 &    --  &    0.277 &    --  &    0.858 &    --  &\bb{0.520}&    --  \\
& Machine QT &    0.355 &  0.004 &    0.223 &  0.006 &    0.768 &  0.035 &    0.460 &  0.062 \\
& Machine DT &    0.411 &  0.549 &    0.260 &  0.489 &\bb{0.863}&  0.866 &    0.476 &  0.319 \\
& mBERT      &    0.324 &  0.009 &    0.179 &  0.004 &    0.768 &  0.035 &    0.314 &  0.000 \\
& infoXLM    &\bb{0.514}&  0.040 &\bb{0.366}&  0.015 &    0.768 &  0.035 &\bb{0.520}&  1.000 \\
& XLM-R      &    0.499 &  0.078 &    0.349 &  0.042 &    0.768 &  0.035 &    0.504 &  0.741 \\

\midrule
\multicolumn{1}{l}{\multirow{3}{*}{Russian}} 
& Human QT   &    0.373 &     -- &    0.239 &     -- &\bb{0.760}&     -- &\bb{0.448}&     -- \\
& Machine QT &    0.335 &  0.237 &    0.217 &  0.386 &    0.710 &  0.154 &    0.366 &  0.285 \\
& Machine DT &    0.348 &  0.533 &    0.213 &  0.424 &    0.756 &  0.923 &    0.402 &  0.324 \\
& mBERT      &    0.199 &  0.000 &    0.087 &  0.000 &    0.710 &  0.154 &    0.156 &  0.000 \\
& infoXLM    &    0.353 &  0.602 &    0.233 &  0.874 &    0.710 &  0.154 &    0.342 &  0.015 \\
& XLM-R      &\bb{0.377}&  0.906 &\bb{0.249}&  0.743 &    0.710 &  0.154 &    0.414 &  0.384 \\
\bottomrule
\end{tabular}
\vspace{-1em}
\end{table}

\jim{For both nDCG and MAP, human query translation tends to provide the most effective results,
usually indistinguishable from machine document translation and from XLM-R (both of which are effective but computationally expensive).
In contrast, machine query translation is efficient.
Title queries are unlikely to be grammatically sound though,
so machine translation quality is lower, resulting in lower retrieval effectiveness.}
We report p-values for two-sided pairwise statistical significance tests. As expected with this number of topics~\cite{clough2013evaluating}, some differences that would be significant at $p<0.05$ are observed.\footnote{Bonferonni correction for 5 tests yields $p<0.01$ for significance.}

\doug{The similar levels of \textit{Judged at 10} (the fraction of the top 10 documents that were judged) among the highest-scoring systems by nDCG and MAP suggest that our relevance judgments are not biased toward any of those systems, despite their diverse designs.} mBERT yields specifically lower \textit{Judged at 10} due to the significantly worse effectiveness, which has also been found by others~\cite{hu2020xtreme}.

\section{Conclusion}

Our new HC4 test collections provide a basis for comparing the retrieval effectiveness of both traditional and neural CLIR techniques. HC4 allows for wide
distribution since documents are distributed as part of the Common Crawl and the topics and relevance judgments are being made freely available for research use.
HC4 is among the first collections in which 
judged documents are 
\doug{principally}
identified using active learning. In addition to providing titles and descriptions
in English and in the language of the documents,
English contemporaneous reports are included to support research into using
additional context for retrieval. HC4 will thus help 
enable development of 
next generation 
CLIR algorithms.

\bibliographystyle{splncs04}
\bibliography{ms}

\begin{thebibliography}{10}
\providecommand{\url}[1]{\texttt{#1}}
\providecommand{\urlprefix}{URL }
\providecommand{\doi}[1]{https://doi.org/#1}

\bibitem{hical}
Abualsaud, M., Ghelani, N., Zhang, H., Smucker, M.D., Cormack, G.V., Grossman,
  M.R.: A system for efficient high-recall retrieval. In: The 41st
  International ACM SIGIR Conference on Research \& Development in Information
  Retrieval. pp. 1317--1320. ACM (2018)

\bibitem{bajaj2016ms}
Bajaj, P., Campos, D., Craswell, N., Deng, L., Gao, J., Liu, X., Majumder, R.,
  McNamara, A., Mitra, B., Nguyen, T., et~al.: {MS MARCO}: A human generated
  machine reading comprehension dataset. arXiv preprint arXiv:1611.09268
  (2016)

\bibitem{baron2016perspectives}
Baron, J., Losey, R., Berman, M.: Perspectives on Predictive Coding: And Other
  Advanced Search Methods for the Legal Practitioner. American Bar Association,
  Section of Litigation (2016),
  \url{https://books.google.com/books?id=TdJ2AQAACAAJ}

\bibitem{bonab2020training}
Bonab, H., Sarwar, S.M., Allan, J.: Training effective neural clir by bridging
  the translation gap. In: Proceedings of the 43rd International ACM SIGIR
  Conference on Research and Development in Information Retrieval. pp. 9--18
  (2020)

\bibitem{DBLP:conf/sigir/BuckleyV04}
Buckley, C., Voorhees, E.M.: Retrieval evaluation with incomplete information.
  In: Sanderson, M., J{\"{a}}rvelin, K., Allan, J., Bruza, P. (eds.) {SIGIR}
  2004: Proceedings of the 27th Annual International {ACM} {SIGIR} Conference
  on Research and Development in Information Retrieval, Sheffield, UK, July
  25-29, 2004. pp. 25--32. {ACM} (2004). \doi{10.1145/1008992.1009000}

\bibitem{chi2020infoxlm}
Chi, Z., Dong, L., Wei, F., Yang, N., Singhal, S., Wang, W., Song, X., Mao,
  X.L., Huang, H., Zhou, M.: Infoxlm: An information-theoretic framework for
  cross-lingual language model pre-training. arXiv preprint arXiv:2007.07834
  (2020)

\bibitem{clough2013evaluating}
Clough, P., Sanderson, M.: Evaluating the performance of information retrieval
  systems using test collections. Information Research  \textbf{18}(2) (2013)

\bibitem{conneau2019unsupervised}
Conneau, A., Khandelwal, K., Goyal, N., Chaudhary, V., Wenzek, G., Guzm{\'a}n,
  F., Grave, E., Ott, M., Zettlemoyer, L., Stoyanov, V.: Unsupervised
  cross-lingual representation learning at scale. arXiv preprint
  arXiv:1911.02116  (2019)

\bibitem{cormack2014evaluation}
Cormack, G.V., Grossman, M.R.: Evaluation of machine-learning protocols for
  technology-assisted review in electronic discovery. In: Proceedings of the
  37th international ACM SIGIR conference on Research \& development in
  information retrieval. pp. 153--162 (2014)

\bibitem{CormackZGASGRG19}
Cormack, G.V., Zhang, H., Ghelani, N., Abualsaud, M., Smucker, M.D., Grossman,
  M.R., Rahbariasl, S., Ghenai, A.: Dynamic sampling meets pooling. In:
  Piwowarski, B., Chevalier, M., Gaussier, {\'{E}}., Maarek, Y., Nie, J.,
  Scholer, F. (eds.) Proceedings of the 42nd International {ACM} {SIGIR}
  Conference on Research and Development in Information Retrieval, {SIGIR}
  2019, Paris, France, July 21-25, 2019. pp. 1217--1220. {ACM} (2019).
  \doi{10.1145/3331184.3331354}

\bibitem{patapsco}
Costello, C., Yang, E., Lawrie, D., Mayfield, J.: Patapasco: A {P}ython
  framework for cross-language information retrieval experiments. In:
  Proceedings of the 44th European Conference on Information Retrieval (ECIR)
  (2022)

\bibitem{DavisD95}
Davis, M.W., Dunning, T.: A {TREC} evaluation of query translation methods for
  multi-lingual text retrieval. In: Harman, D.K. (ed.) Proceedings of The
  Fourth Text REtrieval Conference, {TREC} 1995, Gaithersburg, Maryland, USA,
  November 1-3, 1995. {NIST} Special Publication, vol. 500-236. National
  Institute of Standards and Technology {(NIST)} (1995),
  \url{http://trec.nist.gov/pubs/trec4/papers/nmsu.ps.gz}

\bibitem{devlin2018bert}
Devlin, J., Chang, M.W., Lee, K., Toutanova, K.: Bert: Pre-training of deep
  bidirectional transformers for language understanding. arXiv preprint
  arXiv:1810.04805  (2018)

\bibitem{FerroP09}
Ferro, N., Peters, C.: {CLEF} 2009 ad hoc track overview: {TEL} and {Persian}
  tasks. In: Peters, C., Nunzio, G.M.D., Kurimo, M., Mandl, T., Mostefa, D.,
  Pe{\~{n}}as, A., Roda, G. (eds.) Multilingual Information Access Evaluation
  I. Text Retrieval Experiments, 10th Workshop of the Cross-Language Evaluation
  Forum, {CLEF} 2009, Corfu, Greece, September 30 - October 2, 2009, Revised
  Selected Papers. Lecture Notes in Computer Science, vol.~6241, pp. 13--35.
  Springer (2009). \doi{10.1007/978-3-642-15754-7\_2}

\bibitem{ferro2019information}
Ferro, N., Peters, C.: Information Retrieval Evaluation in a Changing World:
  Lessons Learned from 20 Years of {CLEF}, vol.~41. Springer (2019)

\bibitem{ghalandari2020}
Ghalandari, D.G., Hokamp, C., The~Pham, N., Glover, J., Ifrim, G.: A
  large-scale multi-document summarization dataset from the {Wikipedia} current
  events portal. In: Proceedings of the 58th Annual Meeting of the Association
  for Computational Linguistics (ACL '20). pp. 1302--1308 (2020)

\bibitem{totalrecall2016}
Grossman, M.R., Cormack, G.V., Roegiest, A.: {TREC} 2016 total recall track
  overview. In: Voorhees, E.M., Ellis, A. (eds.) Proceedings of The
  Twenty-Fifth Text REtrieval Conference, {TREC} 2016, Gaithersburg, Maryland,
  USA, November 15-18, 2016. {NIST} Special Publication, vol. 500-321. National
  Institute of Standards and Technology {(NIST)} (2016),
  \url{http://trec.nist.gov/pubs/trec25/papers/Overview-TR.pdf}

\bibitem{hieber2017sockeye}
Hieber, F., Domhan, T., Denkowski, M., Vilar, D., Sokolov, A., Clifton, A.,
  Post, M.: Sockeye: A toolkit for neural machine translation. arXiv preprint
  arXiv:1712.05690  (2017)

\bibitem{hu2020xtreme}
Hu, J., Ruder, S., Siddhant, A., Neubig, G., Firat, O., Johnson, M.: Xtreme: A
  massively multilingual multi-task benchmark for evaluating cross-lingual
  generalisation. In: International Conference on Machine Learning. pp.
  4411--4421. PMLR (2020)

\bibitem{DBLP:conf/sigir/JarvelinK00}
J{\"{a}}rvelin, K., Kek{\"{a}}l{\"{a}}inen, J.: {IR} evaluation methods for
  retrieving highly relevant documents. In: Yannakoudakis, E.J., Belkin, N.J.,
  Ingwersen, P., Leong, M. (eds.) {SIGIR} 2000: Proceedings of the 23rd Annual
  International {ACM} {SIGIR} Conference on Research and Development in
  Information Retrieval, July 24-28, 2000, Athens, Greece. pp. 41--48. {ACM}
  (2000). \doi{10.1145/345508.345545}

\bibitem{DBLP:conf/ntcir/KandoKNEKH99}
Kando, N., Kuriyama, K., Nozue, T., Eguchi, K., Kato, H., Hidaka, S.: Overview
  of {IR} tasks. In: Kando, N. (ed.) Proceedings of the First {NTCIR} Workshop
  on Research in Japanese Text Retrieval and Term Recognition, NTCIR-1, Tokyo,
  Japan, August 30 - September 1, 1999. National Center for Science Information
  Systems {(NACSIS)} (1999),
  \url{http://research.nii.ac.jp/ntcir/workshop/OnlineProceedings/IR-overview.pdf}

\bibitem{clef2019ehealth-tar}
Kanoulas, E., Li, D., Azzopardi, L., Spijker, R.: {CLEF} 2019 technology
  assisted reviews in empirical medicine overview. In: CEUR workshop
  proceedings. vol.~2380 (2019)

\bibitem{landis1977measurement}
Landis, J.R., Koch, G.G.: The measurement of observer agreement for categorical
  data. Biometrics pp. 159--174 (1977)

\bibitem{DBLP:journals/sigir/Lewis95a}
Lewis, D.D.: A sequential algorithm for training text classifiers: Corrigendum
  and additional data. {SIGIR} Forum  \textbf{29}(2),  13--19 (1995).
  \doi{10.1145/219587.219592}

\bibitem{lin2021pretrained}
Lin, J., Nogueira, R., Yates, A.: Pretrained transformers for text ranking:
  {BERT} and beyond. Synthesis Lectures on Human Language Technologies
  \textbf{14}(4),  1--325 (2021)

\bibitem{macavaney2019cedr}
MacAvaney, S., Yates, A., Cohan, A., Goharian, N.: {CEDR}: Contextualized
  embeddings for document ranking. In: Proceedings of the 42nd International
  ACM SIGIR Conference on Research and Development in Information Retrieval.
  pp. 1101--1104 (2019)

\bibitem{MajumderMPBMPMS10}
Majumder, P., Mitra, M., Pal, D., Bandyopadhyay, A., Maiti, S., Pal, S., Modak,
  D., Sanyal, S.: The {FIRE} 2008 evaluation exercise. {ACM} Transactions on
  Asian Language Information Processing  \textbf{9}(3),  10:1--10:24 (2010).
  \doi{10.1145/1838745.1838747}

\bibitem{DBLP:journals/tois/MoffatZ08}
Moffat, A., Zobel, J.: Rank-biased precision for measurement of retrieval
  effectiveness. {ACM} Transactions on Information Systems  \textbf{27}(1),
  2:1--2:27 (2008). \doi{10.1145/1416950.1416952}

\bibitem{nair2020combining}
Nair, S., Galuscakova, P., Oard, D.W.: Combining contextualized and
  non-contextualized query translations to improve {CLIR}. In: Proceedings of
  the 43rd International ACM SIGIR Conference on Research and Development in
  Information Retrieval. pp. 1581--1584 (2020)

\bibitem{colbert-x}
Nair, S., Yang, E., Lawrie, D., Duh, K., McNamee, P., Murray, K., Mayfield, J.,
  Oard, D.: Transfer learning approaches for building cross-language dense
  retrieval models. In: Proceedings of the 44th European Conference on
  Information Retrieval (ECIR) (2022)

\bibitem{oard2013information}
Oard, D.W., Webber, W.: Information retrieval for e-discovery. Information
  Retrieval  \textbf{7}(2-3),  99--237 (2013)

\bibitem{totalrecall2015}
Roegiest, A., Cormack, G.V., Clarke, C.L.A., Grossman, M.R.: {TREC} 2015 total
  recall track overview. In: Voorhees, E.M., Ellis, A. (eds.) Proceedings of
  The Twenty-Fourth Text REtrieval Conference, {TREC} 2015, Gaithersburg,
  Maryland, USA, November 17-20, 2015. {NIST} Special Publication, vol.
  500-319. National Institute of Standards and Technology {(NIST)} (2015),
  \url{https://trec.nist.gov/pubs/trec24/papers/Overview-TR.pdf}

\bibitem{DBLP:journals/ir/SakaiK08}
Sakai, T., Kando, N.: On information retrieval metrics designed for evaluation
  with incomplete relevance assessments. Information Retrieval  \textbf{11}(5),
   447--470 (2008). \doi{10.1007/s10791-008-9059-7}

\bibitem{sakai2021evaluating}
Sakai, T., Oard, D.W., Kando, N.: Evaluating Information Retrieval and Access
  Tasks: NTCIR's Legacy of Research Impact. Springer Nature (2021)

\bibitem{salton1970automatic}
Salton, G.: Automatic processing of foreign language documents. Journal of the
  American Society for Information Science  \textbf{21}(3),  187--194 (1970)

\bibitem{SchaubleS97}
Sch{\"{a}}uble, P., Sheridan, P.: Cross-language information retrieval {(CLIR)}
  track overview. In: Voorhees, E.M., Harman, D.K. (eds.) Proceedings of The
  Sixth Text REtrieval Conference, {TREC} 1997, Gaithersburg, Maryland, USA,
  November 19-21, 1997. {NIST} Special Publication, vol. 500-240, pp. 31--43.
  National Institute of Standards and Technology {(NIST)} (1997),
  \url{http://trec.nist.gov/pubs/trec6/papers/clir\_track\_US.ps}

\bibitem{Smeaton96}
Smeaton, A.F.: Spanish and {Chinese} document retrieval in {TREC-5}. In:
  Voorhees, E.M., Harman, D.K. (eds.) Proceedings of The Fifth Text REtrieval
  Conference, {TREC} 1996, Gaithersburg, Maryland, USA, November 20-22, 1996.
  {NIST} Special Publication, vol. 500-238. National Institute of Standards and
  Technology {(NIST)} (1996),
  \url{http://trec.nist.gov/pubs/trec5/papers/multilingual\_track.ps.gz}

\bibitem{thakur2021beir}
Thakur, N., Reimers, N., R{\"u}ckl{\'e}, A., Srivastava, A., Gurevych, I.:
  {BEIR}: A heterogenous benchmark for zero-shot evaluation of information
  retrieval models. arXiv preprint arXiv:2104.08663  (2021)

\bibitem{DBLP:journals/ipm/Voorhees00a}
Voorhees, E.M.: Variations in relevance judgments and the measurement of
  retrieval effectiveness. Information Processing and Management
  \textbf{36}(5),  697--716 (2000). \doi{10.1016/S0306-4573(00)00010-8}

\bibitem{voorhees2021coopetition}
Voorhees, E.M.: Coopetition in {IR} research. SIGIR Forum  \textbf{54}(2)
  (August 2021)

\bibitem{webber2010effect}
Webber, W., Moffat, A., Zobel, J.: The effect of pooling and evaluation depth
  on metric stability. In: EVIA@ NTCIR. pp. 7--15 (2010)

\bibitem{yang2021cost}
Yang, E., Lewis, D.D., Frieder, O.: On minimizing cost in legal document review
  workflows. In: Proceedings of the 21st ACM Symposium on Document Engineering
  (Aug 2021)

\bibitem{yarmohammadi2019robust}
Yarmohammadi, M., Ma, X., Hisamoto, S., Rahman, M., Wang, Y., Xu, H., Povey,
  D., Koehn, P., Duh, K.: Robust document representations for cross-lingual
  information retrieval in low-resource settings. In: Proceedings of Machine
  Translation Summit XVII Volume 1: Research Track. pp. 12--20 (2019)

\bibitem{DBLP:conf/cikm/YilmazA06}
Yilmaz, E., Aslam, J.A.: Estimating average precision with incomplete and
  imperfect judgments. In: Yu, P.S., Tsotras, V.J., Fox, E.A., Liu, B. (eds.)
  Proceedings of the 2006 {ACM} {CIKM} International Conference on Information
  and Knowledge Management, Arlington, Virginia, USA, November 6-11, 2006. pp.
  102--111. {ACM} (2006). \doi{10.1145/1183614.1183633}

\bibitem{zhang2019improving}
Zhang, R., Westerfield, C., Shim, S., Bingham, G., Fabbri, A.R., Hu, W., Verma,
  N., Radev, D.: Improving low-resource cross-lingual document retrieval by
  reranking with deep bilingual representations. In: Proceedings of the 57th
  Annual Meeting of the Association for Computational Linguistics. pp.
  3173--3179 (2019)

\bibitem{zhang2021mr}
Zhang, X., Ma, X., Shi, P., Lin, J.: Mr. {TyDi}: A multi-lingual benchmark for
  dense retrieval. arXiv preprint arXiv:2108.08787  (2021)

\bibitem{zhao2019weakly}
Zhao, L., Zbib, R., Jiang, Z., Karakos, D., Huang, Z.: Weakly supervised
  attentional model for low resource ad-hoc cross-lingual information
  retrieval. In: Proceedings of the 2nd Workshop on Deep Learning Approaches
  for Low-Resource NLP (DeepLo 2019). pp. 259--264 (2019)

\bibitem{zobel1998reliable}
Zobel, J.: How reliable are the results of large-scale information retrieval
  experiments? In: Proceedings of the 21st Annual International ACM SIGIR
  Conference on Research and Development in Information Retrieval. pp. 307--314
  (1998)

\end{thebibliography}

\end{document}